\documentclass[10pt,amssymb,twocolumn]{revtex4}
\usepackage{hyperref}
\usepackage{amsmath}
\usepackage{graphicx}
\usepackage{amssymb} 
\usepackage{amsfonts} 

\begin{document}

%\twocolumn[

\title{Comments on ``Disproof of Bell's theorem''}
\author{Florin Moldoveanu}
\affiliation{Committee for Philosophy and the Sciences, University of Maryland, College Park, MD 20742 }
\email{fmoldove@gmail.com}

\begin{abstract}
In a series of very interesting papers \cite{JoyDisproof1, JoyReply1, JoyConsolidations1, JoyIllusion1, JoyBell1, JoyBound1, JoyDisproof2}, Joy Christian constructed a counterexample to Bell's theorem. This counterexample does not have the same assumptions as the original Bell's theorem, and therefore it does not represent a genuine disproof in a strict mathematical sense. However, assuming the physical relevance of the new assumptions, the counterexample is shown to be a contextual hidden variable theory. If Bell's theorem's importance is to rule out contextual hidden variable theories obeying relativistic locality, then Joy Christian's counterexample achieves its aim. If however contextual hidden variables theories are not considered genuine physically theories and Bell's theorem's importance stems from its ability to be experimentally confirmed, then Joy Christian's counterexample does not diminish the importance of Bell's theorem. The implications of Joy Christian's counterexample are discussed in the context of information theory.

Version 2 note: Subsequent analysis \cite{Florin1} disproved the mathematical consistency of Joy Christian's model. This paper was based on the assumption of the mathematical validity of the model. Except for the addition of this note, the content of this paper was not modified in any other way.
\end{abstract}

%\pacs{ }

\maketitle

In Ref. \cite{JoyDisproof2}, Joy Christian presents a challenge: a concrete local and realistic model which can reproduce the correlations observed in the EPR-Bohm experiments. The intention is to disproof Bell's theorem by counterexample. This counterexample is supposed to be a pure minimalist mathematical argument free of interpretations and justifications. However, the history behind this result is rich in both content and controversy and to proper understand this counterexample, an examination of the original claim \cite{JoyDisproof1}, critic's replies \cite{Marcin1, Holman1, Grangier1, Tung1}, responses to criticism \cite{JoyReply1, JoyConsolidations1}, and the recent mathematical results into this area \cite{JoyIllusion1, JoyBell1, JoyBound1} is necessary. In the following I will not present an exhaustive review of all those papers, but I will pick what I consider to be the key points.

To begin, the fact that Joy Christian's counterexample is not strictly a mathematical disproof was first noticed by Philippe Grangier \cite{Grangier1} and acknowledged by Joy Christian \cite{JoyReply1}. Joy Christian is using Clifford algebras, certainly a new idea not typically used in the context of Bell's theorem. The justification for it comes from a challenge: why should we constrain ourselves to using only commutative local realistic ``beables''? Staying with commutative ``beables'' necessarily yields Bell's theorem. But is this restriction a genuine physical requirement? 

Consider for example how one can represent rotations in scalar algebra. The standard formulation is not well defined everywhere as illustrated by the known ``gimbal lock'' problem faced by airplane designers \cite{JoyReply1}. Representing rotations as quaternions is a much more natural formalism because in this case there are no artificial singularities introduced by the concrete mathematical representation. Just because quaternions do not commute (or equivalently performing a sequence of two rotations depends on the order of operation) does not make rotations less physical, and generalizing ``beables'' to non-commutative quantities is perfectly justified.

For reference here are the algebraic identities satisfied by Bell's ``beables'' \cite{JoyReply1}:

\begin{eqnarray}
\{-A_a(\lambda)\} + A_a(\lambda) = A_a(\lambda) + \{-A_a(\lambda)\} = 0~  \\
A^{-1}_a (\lambda) A_a(\lambda) = A_a(\lambda) A^{-1}_a (\lambda) = 1~  \\
A_a(\lambda) + \{B_b(\lambda) + C_c(\lambda)\} = \{A_a(\lambda) + B_b(\lambda)\} + C_c(\lambda)~  \\
A_a(\lambda) \{B_b(\lambda) C_c(\lambda)\} = \{A_a(\lambda) B_b(\lambda)\} C_c(\lambda)~  \\
A_a(\lambda) \{B_b(\lambda) + C_c(\lambda)\} = A_a(\lambda) B_b(\lambda) + A_a(\lambda) C_c(\lambda)~  \\
\{B_b(\lambda) + C_c(\lambda)\} A_a(\lambda) = B_b(\lambda) A_a(\lambda) + C_c(\lambda) A_a(\lambda)~  \\
A_a(\lambda) + B_b(\lambda) = B_b(\lambda) + A_a(\lambda)~  \\
A_a(\lambda) B_b(\lambda) = B_b(\lambda) A_a(\lambda)~   
\end{eqnarray}

In his generalization, Joy Christian eliminates the last identity, commutativity law of multiplication, and his ``beables'' are bivectors satisfying the following multiplication rule instead \cite{JoyDisproof2}: 
\begin{equation}
\beta_i \beta_j = -\delta_{ij} - \epsilon_{ijk} \beta_k
\end{equation}

with $\{\beta_i , \beta_j , \beta_k \}$ a fixed bivector basis.

But what is the physical justification for non-commutative ``beables'', besides that it is a possible generalization? Joy Christian arrives at Eq.~9 by into taking into consideration both the actual results of an experiment, and all the counterfactual possibilities \cite{JoyBound1}. The justification comes from Joy Christian's (unusual) interpretation of the definition of completeness by EPR: ``every element of the physical reality must have a counterpart in the physical theory''. Not considering the counterfactuals, amounts in Joy Christian's interpretation to a serious ``topological error'' \cite{JoyBell1, JoyBound1} which forfeits any hope of constructing a local realistic model from the very beginning. Equivalently, Bell's theorem is a trivial topological result due to the incompatibility between discrete outcome result $\{-1, +1\}$ and all possible measurement result outcomes which has the topology of a sphere.

It is worth noting that EPR-Bohm experiment is not the only case where Joy Christian managed to find factorizable models. His methodology for constructing factorizable models passed the tests of Bell's inequality, Hardy-type, and the GHZ theorems \cite{JoyIllusion1}. At this point it is very instructive to see why Joy Christian's model worked in the cases he considered. Joy Christian already did the heavy mathematical lifting required, and he found that it is due to the notion of parallelizability of $S^0$, $S^1$, $S^3$, $S^7$ spheres \cite{JoyBound1}. Parallelizability means that one can always define a smooth flowing motion at the same time in any direction on the manifold. Also the maximum quantum mechanical correlations (Tsirelson's bound) correspond to the maximum torsion allowed on the four spheres when using a torsion connection \cite{JoyBound1}.

After those mathematical developments, the original justification in terms of EPR (counterfactual) completeness required some clarification. Following it strictly would lead to $S^2$ instead of $S^3$. Joy Christian improves on this prescription and replaced it with the following definition: ``completeness = parallelizability'' \cite{JoyBound1}. The new justification is rooted into the closure under multiplication for the four normed division algebras: real numbers, complex numbers, quaternions, and octonions. However, this definition is unsatisfactory for three reasons. First, it is not the parallelizability itself that is physically relevant, but the EPR completeness based on counterfactuals. Second, the cases which Joy Christian considered were exceptional because of the ability to embed the counterfactual space of all potential experimental outcomes in one of the fours spheres. Third, there could be cases where counterfactuals have a topology which would prevent such an embedding and yet one cannot claim in this case that quantum mechanics is ``incomplete''. Upon a successful embedding, the counterfactuals are called ``elements of reality'' by Joy Christian.

Joy Christian then attempts a generalization and makes an extraordinary claim without proof \cite{JoyIllusion1}: for any Hilbert space $\cal{H}$ of arbitrary dimension there exist a topological space $\Omega$ corresponding to EPR counterfactual elements of reality and a morphism  $m: {\cal H}\rightarrow \Omega$ such that the following relation holds:

\begin {equation}
m(|\Psi_a > + |\Psi_b>) = m(|\Psi_a>)m(|\Psi_b>)=A_a A_b \in \Omega 
\end{equation}

with $A_a$ and $A_b$ two points in the topological space $\Omega$. 

From this it is not hard to prove that exact local-realistic completion of any arbitrary entangled state is always possible. At this point, it is not clear if Eq.~10 can always be achieved, and this is a very interesting open problem.  

Let us first note that even if Eq.~10 is not universally valid, this does not impact the validity of the actual counterexamples for EPR-Bohm case, or the Hardy-type or GHZ cases. Also the space $\Omega$ is not simply the topological space of counterfactuals, but a suitable embedding of this space into a parallelized manifold. Upon closer inspection, Eq.~10 (which does hold in the considered cases) suggests that Joy Christian factorizable examples belongs to the realm of contextual theories. The factorization, ``beables'', and hidden variables depends critically on the topology of the space $\Omega$. Preparing many identical copies of the quantum system and then selecting different experiments to be performed may generate different spaces $\Omega$ with different hidden variables. 

Let us try to make this implicit dependency explicit. Joy Christian considered only the EPR-Bohm, Hardy, and GHZ cases. He did not consider another variant of Bell's theorem due to Kochen and Specker \cite{KS1}, Heywood and Redhead \cite{Heywood1}, and Stairs \cite{Stairs1}. This variant is based on the following spin one state:

\begin {equation}
|\Psi> = \frac{1}{\sqrt 3}(|1,-1> - |0,0> + |-1,1>)
\end{equation}

In quick summary, it works like this. Following Peres' approach of proving the Kochen-Specker theorem \cite{PeresBook1}, we know that in this case the results of the experiments are contextual. Measuring spin squared on any three orthogonal direction ($x$, $y$, $z$) always yields two ones and one zero. However it is impossible to assign ahead of time the zero and one values when considering a particular set of 33 directions in space as shown by Peres. Therefore the outcome of the measurement for $s_z^2$ for particle 1 on the $z$ direction is dependent on the $s_x^2$ and $s_y^2$ context. Also measuring $s_z^2$ for particle 2 is perfectly correlated with $s_z^2$ for particle 1. If remote contextual independence is assumed, measuring $s_z^2$ for particle 1 should not depend on the context for particle 2 and therefore one arrives at a contradiction.

For clarity, as a side note, let us also define contextual independence. First consider the following probability: $p^1_m (s|a, b)$ = the probability that the outcome of the measurement performed on 1 is s when m is the complete state, the measurements performed on 1 and 2 respectively are a and b.

Contextual independence is defined as follows \cite{Shimony0}:

\begin{eqnarray}
p^1_m (s|a, b) &=& p^1_m (s| a) \rm{~independent~of~b}\\
p^2_m (t|a, b) &=& p^2_m (t| b) \rm{~independent~of~a}
\end{eqnarray}

For commutative beables factorization is equivalent with contextual independence and outcome independence \cite{Jarett1}. Joy Christian claims that his counterexample satisfies both contextual and outcome independence \cite{JoyDisproof2} because it satisfies factorization. However, this claim is incorrect as it will be shown below.

Ignoring contextual independence, let us apply Joy Christian's method to the spin 1 case of Eq.~11. Can we find in this case a morphism obeying Eq.~10? This is a hard problem, but fortunately we do not need to solve it and we can find directly the hidden variables. We know that polarization entangled four-photon fields (2-photons in each of two spatial modes) of pulsed parametric down-conversion are formally equivalent to two maximally entangled spin-1 particles as shown in Ref.~\cite{Howell1}. Then we can apply Joy Christian's model for each spin one half, and construct explicitly the local hidden variables (which are the two local hidden variables of the two pairs of spin one half systems).  Now due to Kochen-Specker theorem, no context free combining of the two local hidden spin one half variables is possible in general and therefore Joy Christian's spin one hidden variables must be contextual. In turn this implies that the spin one half hidden variables are contextual as well. It is now clear how noncontextuality is violated by Joy Christian's model: within any given context $\Omega$, Eqs.~12 and 13 are satisfied (for example in the original $S^3$ space of spin one half system), but changing the space $\Omega$ by performing different experiments makes them invalid. Eqs.~12 and 13 should be valid in general for all conceivable experiments in order to be able to claim contextual independence. Shimony coined a name for such a contextual hidden variable theory, calling it ``environmental'' \cite{Shimony1}. The space $\Omega$ for spin 1 is not one of the four parallelizable spheres (because if it would have been then we could have had a unique assignment of experimental outcomes in opposition to Kochen-Specker theorem) and the morphism of Eq.~10 if it indeed exist it would be definable only on local neighborhoods similar with how a manifold requires in general an atlas of maps to a local ${\mathbb{R}}^n$. 

Other people considered the possibility of factorization for contextual hidden variable models \cite{Shimony1}\cite{Shafiee1}\cite{Shafiee2}. In Ref. \cite{Shimony1}, Shimony states:

{\it ``It has been suggested by Fine (private correspondence) that joint probability distributions of non-commuting observables can be introduced in contextual hidden variables theories by speaking counter-factually''.}

This certainly anticipates Joy Christian's approach. Also Shafiee produced a contextual stochastic hidden variable theory which cannot be ruled out by Bell's theorem \cite{Shafiee2}.

So what does this mean for a ``disproof'' of Bell's theorem? To answer this we need to discuss the importance attached to Bell's theorem. If Joy Christian's counterexample does not strictly follow from Bell's original assumptions of commuting local ``beables'', can it be considered a ``disproof'' in terms of its importance into eliminating hidden variable theories? Hidden variable theories belong to two broad classes: non-contextual and contextual. Non-contextual hidden variable theories are excluded by Gleason's or Kochen-Specker's theorems, but they cannot be directly experimentally tested. On the other hand, (contextual and non-contextual) realistic hidden variable theories subject to a certain locality condition are ruled out by Bell's theorem, and experiments do confirm the violations of Bell inequalities precisely as predicted by quantum mechanics.

Quoting Shimony \cite{Shimony1}: {\it ``The world view associated with noncontextual hidden variables theories is overwhelmingly discontinued, whether or not one pays attention to relativistic locality. That associated with contextual hidden variables theories is overwhelmingly disconfirmed if relativistic locality is assumed. Bell's Theorem is a useful instrument for strengthening the first of these judgments, and at present it is indispensable for achieving the latter.''}

By providing an explicit factorizable model, Joy Christian's example only disproves the importance of Bell's theorem as an argument against contextual hidden variable theories.  Bell's theorem retains its full importance as the key result which suggested experimental tests for settling the hidden variable theory questions. However, contextual hidden variable theories have not secured the status of physical theories \cite{BCBook1}. Replacing the counterintuitive quantum mechanics with a contextual hidden variable theory does not achieve much in terms of clarity. It is more fitting to insist on strong contextual independence for any experimental settings for a hidden variable theory and this invalidates the importance of Joy Christian's counterexample. 

Last, let us investigate the usage of non-commutative ``beables'' form the point of view of classical and quantum information theory. Classical bits, as the fundamental unit of information should have no internal degrees of freedom and insisting on commutativity is fully justified. In this context Bell's theorem is safe and all modern advances in this area are not affected by Joy Christian's counterexample. If rotations can be used as ``actionable beables'' to transmit classical signals, then classical violations of Bell's inequality would be possible. One can imagine using actual gears similar with the ones found in clocks to implement rotations as a classical resource. However relativity forbids rigid bodies and non-commutative ``beables'' cannot be used as a valid resource to achieve Tsirelson's bound in a classical setting.

Can we make the above comments precise and prove that non-commutative ``beables'' cannot be used in practice to achieve Tsirelson's bound? Indeed we can and the key result is a paper by Clifton \cite{Clifton1} which proved all relevant mathematical theorems needed in this case. The first problem with ``beables'' is their fuzzy meaning, but we can follow Bell in expecting that not all observables are ``beables'', and observables are made out of ``beables''. Clifton then defines precisely the algebra of ``beables'' following Segal's common axiomatization of classical and quantum mechanics. To this aim he calls the algebra of ``beables'' a ``Segalgebra'', or in modern parlance, a Jordan-Lie-Banach algebra (the self-adjoined part of a $C*$ algebra). Clifton then proves two theorems which demand the cases where ``beables'' must be commutative. 

First, commutativity is required whenever a subalgebra of a Segalgebra $S$ has ``beable'' status for every state in a full set of states on $S$. Clearly, Joy Christian's counterexample violates the requirements for this theorem because his counterexample is an ``environmental'' hidden variable theory. 

Second, the following key theorem holds: {\it ``Subalgebras of local ``beables'' selected from the Segalgebras of local observables in relativistic quantum field theory must be commutative in any state of the field with bounded energy.''}

The second theorem clearly shows that Joy Christian's counterexample cannot be used to achieve Tsirelson's bound with classical resources only. To be fully compatible with relativity, ``beables'' must be commuting, and in this case Bell's theorem does hold.

\section*{Acknowledgment}
I want to thank Jeffrey Bub for useful discussions.


\begin{thebibliography}{99}
\bibitem{JoyDisproof1}``Disproof of Bell's Theorem by Clifford Algebra Valued Local Variables'', Joy Christian, arXiv:quant-ph/0703179
\bibitem{JoyReply1}``Disproof of Bell's Theorem: Reply to Critics'', Joy Christian,arXiv:quant-ph/0703244
\bibitem{JoyConsolidations1}``Disproof of Bell's Theorem: Further Consolidations'', Joy Christian, arXiv:quant-ph/0707.1333
\bibitem{JoyIllusion1}``Disproofs of Bell, GHZ, and Hardy Type Theorems and the Illusion of Entanglement'', arXiv:quant-ph/0904.4259
\bibitem{JoyBell1}``Can Bell's Prescription for Physical Reality Be Considered Complete?'', Joy Christian, arXiv:quant-ph/0806.3078
\bibitem{JoyBound1}``What Really Sets the Upper Bound on Quantum Correlations?'', Joy Christian, arXiv:quant-ph:1101.1958
\bibitem{JoyDisproof2}``Disproof of Bell's Theorem'', Joy Christian, arXiv:quant-ph/1103.1879
\bibitem{Marcin1}``Comment on: Disproof of Bell's Theorem by Clifford Algebra Valued Local Variables'', Marcin Pawlowski, arXiv:quant-ph/0703218
\bibitem{Holman1}``Non-Viability of a Counter-Argument to Bell's Theorem'', Marc Holman, arXiv:quant-ph/0704.2038
\bibitem{Grangier1}````Disproof of Bell's Theorem'' : more critics'', Philippe Grangier, arXiv:quant-ph/0707.2223
\bibitem{Tung1}``Disproof of ``Disproof of Bell's Theorem by Clifford Algebra Valued Local Variables'''', Tung Ten Yong, arXiv:quant-ph/0712.1637
\bibitem{KS1} S.~Kochen and E.~P.~Specker, J.~Math.~Mech. \textbf{17}, 59 (1967).
\bibitem{Heywood1} P.~Heywood and M.~L.~G.~Redhead, Found.~Phys. \textbf{13}, 481 (1983). 
\bibitem{Stairs1} A.~Stairs, Phil.~Sci. \textbf{50}, 578 (1983).
\bibitem{PeresBook1} A.~Peres, \emph{Quantum Theory: Concepts and Methods}, Kluwer (1993).
\bibitem{Shimony0} ``Bell's Theorem'' A.~Shimony, http://plato.stanford.edu/entries/bell-theorem/
\bibitem{Jarett1} J.~P.~Jarrett ``On the physical significance of the locality conditions in the Bell arguments'' No\^us \textbf{18}, 569 (1984).
\bibitem{Howell1} J.~C.~ Howell, A,~Lamas-Linares, D.~Bouwmeester, Phys.~Rev.~Lett. \textbf{88}, 030401 (2002).  arXiv:quant-ph/0105132v2
\bibitem{Shimony1} A.~Shimony, Brit.~J.~Phil.~Sci. \textbf{35} 24 (1984).
\bibitem{Shafiee1} ``The Possibility of Factorizable Contextual Hidden Variable Theories'', A.~Shafiee and M~Golshani, arXiv:quant-ph/9907060v2
\bibitem{Shafiee2} ``Common Cause and Contextual Realization of Bell Correlation'', A~Shafiee, R.~Maleeh, and M.~Golshani, arXiv:quant-ph/0504121v1
\bibitem{BCBook1} E.~G.~Beltrametti-G.~Cassinelli \emph{The logic of quantum mechanics} Addison- Wesley Publishing. Co., Reading, Mass. (1981).
\bibitem{Clifton1} ``Beables in algebraic quantum mechanics'', R.~Clifton, arXiv:quant-ph/9711009v1
\bibitem{Florin1} ``Disproof of Joy Christian's ``Disproof of Bell's theorem''
'', F.~Moldoveanu, arXiv:quant-ph/1109.0535v1

\end{thebibliography}
\end{document}